\documentclass[prc, 10pt, twocolumn]{revtex4-1}

\usepackage[utf8]{inputenc}
\usepackage{amsmath}
\usepackage{amssymb}
\usepackage{amsfonts}
\usepackage{newtxtext,newtxmath}
\usepackage{bbold}
\usepackage{bm}
\usepackage{graphicx}
\usepackage[usenames,dvipsnames]{xcolor}
\usepackage{color}
\usepackage[colorlinks=true,linkcolor=blue,urlcolor=blue,citecolor=blue]{hyperref}

\usepackage{slashed}
\usepackage[english]{babel}
\usepackage{dcolumn}
\usepackage{blindtext}
\usepackage{epsfig}
\usepackage{pifont}
\usepackage{dsfont,mathrsfs}
\usepackage{cancel}
\usepackage{bigints}
\usepackage{accents}
\usepackage{soul}
\usepackage{multirow}
\usepackage[caption = false]{subfig}

\begin{document}
\title{Centrality dependence of chemical freeze-out parameters and strangeness equilibration in RHIC and LHC energies}
\author{Deeptak Biswas}
\email{deeptak@jcbose.ac.in}

\affiliation{
Department of Physics, Center for Astroparticle Physics \& Space Science\\
Bose Institute, EN-80, Sector-5, Bidhan Nagar, Kolkata-700091, India 
}
%
%
\begin{abstract}
We have estimated centrality variation of chemical freeze-out parameters from yield data at mid-rapidity of $\pi^\pm$, $K^\pm$ and $p$, $\bar{p}$ for collision energies of RHIC (Relativistic Heavy Ion Collider), Beam Energy Scan (RHIC-BES) program, and LHC (Large Hadron Collider). We have considered a simple hadron resonance gas model and employed a formalism involving conserved charges ($B, Q, S$) of QCD for parameterization. Along with temperature and three chemical potentials ($T,\mu_B,\mu_Q,\mu_S$), a strangeness under-saturation factor ($\gamma_S$) has been used to incorporate the partial equilibration in the strange sector. Our obtained freeze-out temperature does not vary much with centrality, whereas chemical potentials and $\gamma_S$ seem to have a significant dependence. The strange hadrons are found to deviate from a complete chemical equilibrium at freeze-out at the peripheral collisions. This deviation appears to be more prominent as the collision energy decreases at lower RHIC-BES energies. We have also shown that this departure from equilibrium reduces towards central collisions, and strange particle equilibration may happen after a threshold number of participants in $A$-$A$ collision.    
\end{abstract}

\keywords{
Heavy Ion collision, Centrality, Chemical freeze-out, Hadron Resonance Gas model
}
\pacs{12.38.Mh,  21.65.Mn, 24.10.Pa, 25.75.−q}
\maketitle

\section{\label{sec:Intro} Introduction}
In the last few decades, several ion collider experiment collaborations have been developed to explore the phase diagram of quantum chromodynamics (QCD). Relativistic Heavy Ion Collider (RHIC) started to investigate the signatures of deconfined quark-gluon plasma, whereas the RHIC Beam Energy Scan (BES) program became motivated in searching the QCD critical point \cite{Abelev:2009bw}. The Large Hadron Collider (LHC) is trying to investigate the medium created in zero baryon density, where a crossover transition from a hadronic state to a deconfined state of quarks and gluons may happen \cite{FOKA2016154}.

The hot and dense fireball created in these collisions experiences a fast expansion due to the initial pressure gradient. Assuming that the system starts from a strongly interacting quark-gluon state, a fast thermalization can occur. The thermodynamic parameters like temperature and chemical potentials can describe this thermalized medium. The matter and energy density dilute with the expansion and the temperature declines. As the energy density(temperature) drops below the hadronization threshold, the matter evolves as a state of hadrons and their resonances. The mean free path increases with further expansion and various collisions among particles abate. In this context, one can define freeze-out as the boundary, onwards which no interaction is supposed to happen among hadrons.  In the standard description, two freeze-out surfaces are described, depending on the interaction type. The chemical freeze-out(CFO) happens when inelastic scattering stops and the particle abundances become fixed. The kinetic freeze-out (KFO) is the point where elastic collisions cease. In this free non-interacting limit, the ideal hadron resonance gas model may give a reasonable description of the hadrons at freeze-out.

Yields of strange hadrons help to understand the extent of chemical equilibrium achieved in these collisions. The strange quark equilibrates later than the $u$, $d$ quarks due to its larger mass \cite{Koch:1986ud}. Equilibrated strangeness spectra may be a suitable signature to understand the existence of a deconfined partonic phase \cite{PhysRevLett.48.1066}. Recently, a charged particle multiplicity ($dN_{ch}/d\eta$) dependent production of strangeness has been observed in LHC \cite{ALICE:2017jyt}. This strangeness production is related to the collision centrality and number of participants ($N_{part}$) of the collision system \cite{Abelev:2008ab, Aamodt:2010cz}. Here, $N_{part}$ denotes the average number of participating nucleons of a particular collision system.

Following the success of the Statistical Hadronization Model (SHM) \cite{Cleymans:1998fq}, studies have determined the chemical freeze-out parameters, considering the Hadron Resonance Gas (HRG) model. In this context, a $\chi^2$ fitting with the available yield data is well-practiced \cite{Cleymans:2004hj, Cleymans:2005xv, Andronic:2005yp, Andronic:2008gu, Manninen:2008mg, Andronic:2012ut, Chatterjee:2013yga, Alba:2014eba, Chatterjee:2015fua, Adak:2016jtk, Chatterjee:2016cog, Chatterjee:2017yhp}. Generally, one extracts the chemical freeze-out temperature $T_{CFO}$ and the baryon chemical potential $\mu_B$ by the minimization procedure, whereas the charge chemical potential $\mu_Q$ and the strange chemical potential $\mu_S$ get fixed from the constraints of the colliding nuclei. To scale the possible non-equilibration of strange hadrons, a  strangeness under saturation factor $\gamma_S$ can be introduced \cite{Rafelski:1991rh, Letessier:1993hi, Cleymans:1997ib, Hamieh:2000tk, Cleymans:2001at, Bearden:2002ib, Tawfik:2014dha, Manninen:2008mg}. This parameter scales the deviation of strange hadrons from a complete equilibrium in the Grand Canonical Ensemble (GCE). In a recent work \cite{Bhattacharyya:2019cer}, we have shown that in $\chi^2$ analysis, a  larger systematic variation can arise depending on the chosen set of ratios. A conserved-charge-dependent extraction of thermal parameters has been proposed~\cite{Bhattacharyya:2019wag}, which seems to suitably estimate thermal parameters and predict equilibration in the most central collision. It will be interesting to check the centrality variation of thermal parameters and the equilibration of strange particles in this framework.

In this manuscript, we have tried to study the centrality variation of freeze-out parameters, with an emphasis on the saturation of strangeness equilibration~\cite{ALICE:2017jyt} in heavy nuclei collision. We have observed a similar saturation with centrality ($N_{part}$) for collision energies ranging from 7.7 GeV of RHIC-BES to LHC (2.76 TeV). We have employed a recently developed parameter extraction process~\cite{Bhattacharyya:2019wag, Biswas:2020kpu} with a strangeness suppression factor $\gamma_S$ to measure the possible deviation of strange hadrons from respective equilibrium yield. We have found the kaons to deviate from equilibrium at chemical freeze-out for the peripheral and semi-peripheral collision, though the temperature does not change much with centrality. We have further studied the scaling behavior of all freeze-out parameters. The parameters attain a saturation onwards $N_{part}=150$. This flattening indicates that the system created in the heavy-ion collision reaches a grand canonical limit corresponding to the most central value, in which thermodynamic description becomes independent of the system size. Finally, we have verified the efficacy of our parametrization by compared our estimated hadron yield ratios with experimental data.     

We have organized the manuscript as follows. A short description of the Hadron resonance gas model (HRG) is given in section \ref{sec:HRG}. In section \ref{sec:scheme}, we have briefly discussed the parameter extraction method in our approach. \ref{sec:result} describes our results followed by discussion. We summarise our results in section \ref{sec:conclusion}.

\section{\label{sec:HRG}Hadron resonance gas model}
The hadron resonance gas (HRG) model describes the system as a mixture of hadrons and their resonances. It is a standard exercise to incorporate all available hadron yields for obtaining a good description of the medium. In recent years various studies have been performed using HRG model \cite{Hagedorn:1980kb, Rischke:1991ke, Cleymans:1992jz, BraunMunzinger:1994xr, Cleymans:1996cd, Yen:1997rv, BraunMunzinger:1999qy, Cleymans:1999st, BraunMunzinger:2001ip, BraunMunzinger:2003zd, Karsch:2003zq, Tawfik:2004sw, Becattini:2005xt, Andronic:2005yp, Andronic:2008gu, Begun:2012rf, Andronic:2012ut, Tiwari:2011km, Fu:2013gga, Tawfik:2013eua, Garg:2013ata,Bhattacharyya:2013oya, Kadam:2015xsa, Kadam:2015fza, Kadam:2015dda, Albright:2014gva, Albright:2015uua, Begun:2016cva, Sharma:2018jqf, Vovchenko:2019kes}. This model has successfully described hadron yields from AGS to LHC energies \cite{BraunMunzinger:1994xr, Cleymans:1996cd, BraunMunzinger:1999qy, Cleymans:1999st, BraunMunzinger:2001ip, Becattini:2005xt, Andronic:2005yp, Andronic:2008gu}. Bulk properties of hadronic matter have also been studied in this model \cite{Karsch:2003zq, Tawfik:2004sw, Andronic:2012ut}.
 
In the present work, we have considered the ideal HRG model in which hadrons are treated as point-like particles. A grand canonical ensemble can describe the partition function of hadron resonance gas as \cite{Andronic:2012ut},
\begin {equation}
\ln Z^{ideal}=\sum_i \ln Z_i^{ideal},
\end{equation}
The sum runs over all hadrons and resonances. In the idealistic scenario of a chemical freeze-out, we can neglect all dissipative interactions and finite volume corrections. The thermodynamic potential for $i$'th species is given as,
\begin{equation}
\ln Z_i^{ideal}=\pm \frac{Vg_i}{(2\pi)^3}\int d^3p
\ln[1\pm\exp(-(E_i-\mu_i)/T)],
\end{equation}
where the upper sign is for baryons and lower for mesons. Here $V$ is the volume and $T$ is the temperature of the system. For the $i^{th}$ species of hadron, $g_i$, $E_i$ and $m_i$ are respectively the degeneracy factor, energy, and mass, while $\mu_i=B_i\mu_B+S_i\mu_S+Q_i\mu_Q$ is the chemical potential, with $B_i$, $S_i$ and $Q_i$ denoting the baryon number, strangeness and
the electric charge respectively. For a thermalized system, the number density $n_i$ can be calculated from partition function as,
\begin{equation}{\label{density}}
 n_i =\frac{T}{V} \left(\frac{\partial \ln Z_i}
       {\partial\mu_i}\right)_{V,T} 
 =\frac{g_i}{{(2\pi)}^3} \int\frac{d^3p} {\exp[(E_i-\mu_i)/T]\pm1}.
\end{equation}
\section{\label{sec:scheme}Application to Freeze-out}
We first outline the usual application of the HRG model for
characterizing the freeze-out temperature and chemical potentials in the
context of heavy-ion collision experiments. The rapidity density for
$i$'th hadron may be related to the corresponding number density as
\cite{Manninen:2008mg},
\begin{equation}{\label{eq.solve}}
\frac{dN_i}{dy}|_{Det}={\frac{dV}{dy}}n_i^{Tot}|_{Det}
\end{equation}
where the subscript $Det$ denotes the detected hadrons. Here the total number density of any hadron is,
\begin{eqnarray}
&n_i^{Tot}& ~=~ n_i(T,\mu_B,\mu_Q,\mu_S)~ + 
\nonumber \\
&\sum_j& n_j(T,\mu_B,\mu_Q,\mu_S) \times Branch~Ratio (j
\rightarrow i)
\end{eqnarray}
The summation is over the heavier resonances ($j$) that decay to the
$i^{th}$ hadron. This number density $n_i$ is calculated using Eq.\ref{density}.

In this context, it is also important to consider the constraints regarding the conserved charges. Following the assumption of an isentropic evolution, one can employ conservation conditions like strangeness neutrality (Eq.\ref{eq:ns}) and baryon density to charge density (Eq.\ref{eq:nbq}) to restrict the values of chemical potentials \cite{Alba:2014eba}. 
\begin{equation}
\label{eq:ns}
\sum_i n_i (T, \mu_B, \mu_S, \mu_Q) S_i=0,
\end{equation}
\begin{equation}
\label{eq:nbq}
 \sum_i n_i (T, \mu_B, \mu_S, \mu_Q) Q_i= r \sum_i n_i (T, \mu_B, 
   \mu_S, \mu_Q) B_i,
\end{equation}
here, $r$ is the net-charge to net-baryon number ratio of the colliding
nuclei. For example, in Au + Au collisions $r = N_p /(N_p + N_n)=0.4$,
with $N_p$ and $N_n$ denoting the number of protons and neutrons in the
colliding nuclei. In a proton-proton collision, this ratio is $1$.

The usual approach should be solving Eq.\ref{eq.solve} to extract thermal parameters. The freeze-out description will be more reasonable if we include data for a larger number of detected particles in our solving mechanism. So a $\chi^2$ minimization is performed with all the available yields. One may avoid the volume systematics by taking ratios of two hadrons. In this approach, the effects of hydrodynamical flow also disappear \cite{Cleymans:1999st}. Further, performing a minimization procedure with available yield ratios, one can parameterize the chemical freeze-out surface. In a recent work Ref.\cite{Bhattacharyya:2019cer}, we have shown that significant systematic uncertainty may arise in $\chi^2$ analysis due to variation in the chosen set of ratios.
 
\subsection*{\label{sec:our_scheme}Our approach}
Following the complication regarding the chosen set of ratios, we have introduced an alternative method in Ref.\cite{Bhattacharyya:2019wag}. The individual hadrons are not a conserved quantity in the strong interaction. So we opted to introduce ratios regarding conserved net charge densities like $B, Q, S$. Along with the constraints 
Eq.~(\ref{eq:ns}$-$\ref{eq:nbq}), we have proposed two new independent equations, the net baryon number normalized to the total baryon number and the net baryon number normalized to the total hadron yield\cite{Biswas:2019wtp}, as given below.
\begin{eqnarray}
\frac{\sum_i^{Det} B_i \frac{dN_i}{dY}}{\sum_i^{Det} |B_i|
\frac{dN_i}{dY}}
&=& \frac{\sum_i^{Det} B_i n_i^{Tot}}{\sum_i^{Det} |B_i| n_i^{Tot}} 
\label{eq.conserveb} \\
\frac{\sum_i^{Det} B_i \frac{dN_i}{dY}}
{\sum_i^{Det}\frac{dN_i}{dY}}
&=& \frac{\sum_i^{Det} B_i
n_i^{Tot}}{\sum_i^{Det} n_i^{Tot}} 
\label{eq.t}
\end{eqnarray}
We want to mention that the left-hand side consists of the particle yields data from the heavy-ion collision, and those on the right are the number densities calculated from the thermal HRG model. $i$ runs only over detected($Det$) hadrons with available experimental yield data. 
  
\subsection*{\label{sec:centrality}Application to centrality}
The geometric information of the collision system is crucial to understand, as different final observables like eccentricity, elliptical flow, charge particle multiplicity are directly dependent on the initial conditions like impact parameter ($b$), the number of participating nucleons ({$N_{part}$}) \cite{Miller:2007ri}. We employ the centrality bins to differentiate collision events according to their impact parameters. As there is no direct method to measure $b$, the centrality bins can be calibrated from charge particle multiplicity with the Glauber model~\cite{Miller:2007ri, Abelev:2008ab,  Abelev:2013qoq}. Each centrality bin is represented by a corresponding $N_{part}$. Most central ($0\%-5\%$ centrality) collisions correspond to the events with the lowest value of impact parameters (Highest value of $N_{part}$) whereas, the most peripheral ($70\%-80\%$) are with the largest impact parameter and smallest $N_{part}$. The degree of equilibration of the created medium should strongly depend on centralities as the system's initial volume and initial energy, baryon deposition depends on these initial specifications.
    
\subsection*{\label{sec:strange_centrality}Strangeness with centrality}
Introducing a strangeness suppression factor is optional for the most central collisions \cite{Bhattacharyya:2019wag, Bhattacharyya:2019cer}, whereas this appears essential when we deal with a peripheral or semi-central collision. Complete chemical equilibrium may not be achieved in the strangeness sector due to the higher mass threshold of strange particles and their hadronic counterpart \cite{Koch:1986ud}. Initially, this suppression factor $\gamma_S$ was introduced considering the phase space under-saturation \cite{Letessier:1993hi}. Ref.~[\cite{Cleymans:1997ib, Hamieh:2000tk}] has discussed this under-saturation as an effect of the canonical ensemble consideration of strangeness, where exact strangeness conservation should be considered for a smaller collision system.  In ref.\cite{Becattini:2008ya} a core corona dependent model has also tried to discuss this suppression of strangeness. Irrespective of the reason for this undersaturation, considering this factor $\gamma_S$, gives rise to a better agreement to the thermal description of heavy-ion data.  It seems that the strange sector may have a deviation from the respective grand canonical picture, and this parameter is a measure of that departure \cite{Rafelski:1991rh, Letessier:1993hi, Cleymans:1997ib, Hamieh:2000tk, Cleymans:2001at, Bearden:2002ib, Tawfik:2014dha, Manninen:2008mg}. In the presence of this factor, the number density is modified in the following manner \cite{Manninen:2008mg},
\begin{equation}{\label{new_density}}
 n_i = \frac{g_i}{{(2\pi)}^3} \int\frac{d^3p} {{\gamma_S}^{-n_i^s}\exp[(E_i-\mu_i)/T]\pm1}.
\end{equation}
Here, $n_i^s$ denotes the number of valence strange quarks or anti-quarks in the $i$'th hadron. In this work, we have calculated the number densities following \ref{new_density}. $\gamma_{S}=1$ for all non-strange particles. A smaller value of $\gamma_S$ denotes a larger deviation from the grand canonical limit of equilibrium.  

As we have introduced one added parameter $\gamma_S$, an extra equation is needed to close our system of equations. This parameter is not related to any conserved quantity, rather it is used to describe the possible non-equilibrium of the strange sector. Keeping in mind that we have used only yields of kaons among the strange particles, we have utilized kaon to pion ratio to evaluate the value of $\gamma_S$ in Eq.\ref{gammas}. 
\begin{equation}{\label{gammas}}
\sum_i \frac{ (\frac{K}{\pi})^i_{expt}-(\frac{K}{\pi})^i_{model}}{(\frac{K}{\pi})^i_{model}} = 0
\end{equation}
Here $i$ stands for two possible charges,  i.e $({K}/{\pi})^+=K^+/\pi^+$ and $({K}/{\pi})^-=K^-/\pi^-$. Here we want to reiterate that, for smaller system size (peripheral collisions) the exact strangeness conservation demands the canonical treatment. To study the systematic variation with centrality, we have approached within a Grand Canonical Ensemble (GCE) with the $\gamma_S$ to scale the possible deviation from equilibrium picture. This exercise is well practised in the context of freeze-out parameter extraction for various centrality\cite{Becattini:1997ii, Cleymans:2001at, Cleymans:2002xu, Bearden:2002ib, Becattini:2003wp, Cleymans:2003yp, Tawfik:2014dha}. Finally, we have solved all these five equations Eq. [\ref{eq:ns}$-$\ref{eq.t}] and Eq.[\ref{gammas}] to extract the five parameters (T, $\mu_B$, $\mu_Q$, $\mu_S$ and $\gamma_S$).

 These three quantities are independent so as the constructed total charges. So the correlated uncertainties arising from repeated entries of a single yield (addressed in ref.\cite{Adamczyk:2017iwn}), are reduced in this formalism.

In this analysis, we have used yield data of $\pi^\pm$, $K^\pm$, and $p, \bar{p}$. In this context, $netB$ can be constructed out of net-proton, whereas net charge is the sum of net-pion, net-kaon, and net-proton. This consideration is in line with the general approximation of taking the net proton as a proxy for net baryon number \cite{Hatta:2003wn}. For these set of particles, the above-mentioned equations will be, 
\begin{equation*}
\frac{\frac{dN_p}{dY}-\frac{dN_{\bar{p}}}{dY}}{\frac{dN_p}{dY}+\frac{dN_{\bar{p}}}{dY}}
= \frac{n_p^{Tot}-n_{\bar{p}}^{Tot}}{n_p^{Tot}+n_{\bar{p}}^{Tot}} 
\end{equation*}
\begin{eqnarray*}
\frac{\frac{dN_p}{dY}-\frac{dN_{\bar{p}}}{dY}}
{\frac{dN_{\pi^+}}{dY}+\frac{dN_{\pi^-}}{dY}+\frac{dN_{K^+}}{dY}+\frac{dN_{K^-}}{dY}+\frac{dN_p}{dY}+\frac{dN_{\bar{p}}}{dY}} \\
= \frac{n_p^{Tot}-n_{\bar{p}}^{Tot}}{n_{\pi^+}^{Tot}+n_{\pi^-}^{Tot}+n_{K^+}^{Tot}+n_{K^-}^{Tot}+n_p^{Tot}+n_{\bar{p}}^{Tot}} 
\end{eqnarray*}
Here the $n_i^{Tot}$ denotes the total number density of $i'th$ particles, considering the relevant decay channels. We have considered all the strong decay channels from higher mass resonances, whereas weak decay corrections have been performed depending on the experimental specification \cite{Abelev:2008ab, Abelev:2013vea, Adamczyk:2017iwn}. In LHC, we have not included weak decay contribution into protons, whereas in RHIC energies, they are present in the total density.

\section{\label{sec:result}Results and discussions}
In this analysis we have used the yields of, $\pi^\pm$ (139.57 MeV), $K^{\pm}$ (493.68 MeV) and $p$,$\bar{p}$ (938.27 MeV) for collision energies ($\sqrt{s_{NN}}$) ranging from RHIC-BES (7.7 GeV) to LHC (2.76 TeV). For convenience, we have represented the centrality bins by their corresponding number of participants ($N_{part}$). The collision system is Au-Au at higher RHIC and RHIC-BES energies, and Pb-Pb at LHC. Data have been used following RHIC~\cite{Abelev:2008ab}, RHIC-BES~\cite{Adamczyk:2017iwn, Adam:2019dkq} and LHC~\cite{Abelev:2013vea}. Data for p-p collision is available in RHIC for $\sqrt{s_{NN}}=200$ GeV and included in our analysis for completeness. In the present analysis, we have only taken mid-rapidity data. The details of the experimental yields used in the analysis are listed in the Ref.\cite{Abelev:2008ab, Abelev:2013vea, Adamczyk:2017iwn, Adam:2019dkq}.

In our HRG spectrum, we have used all confirmed hadrons up to 2 GeV, with masses and branching ratios following the Particle Data Group~\cite{Tanabashi:2018oca} and THERMUS \cite{Wheaton:2004qb}, which is a numerical thermal model package for the root framework. Finally, we solve Eq.(\ref{eq:ns}$-$\ref{eq.t}, \ref{gammas}) numerically, using Broyden's method with a minimum convergence criterion of $10^{-6}$ \cite{Press:2007:NRE:1403886}. We have estimated the variances of thermal parameters by repeating the analysis at the given extremum value of hadrons yields.
 
\subsection{Freeze-out Parameters}
\label{sc.freeze}
\begin{figure*}[hbt!] 
\subfloat[]{
\hspace{-2em}
{\includegraphics[width=\textwidth]{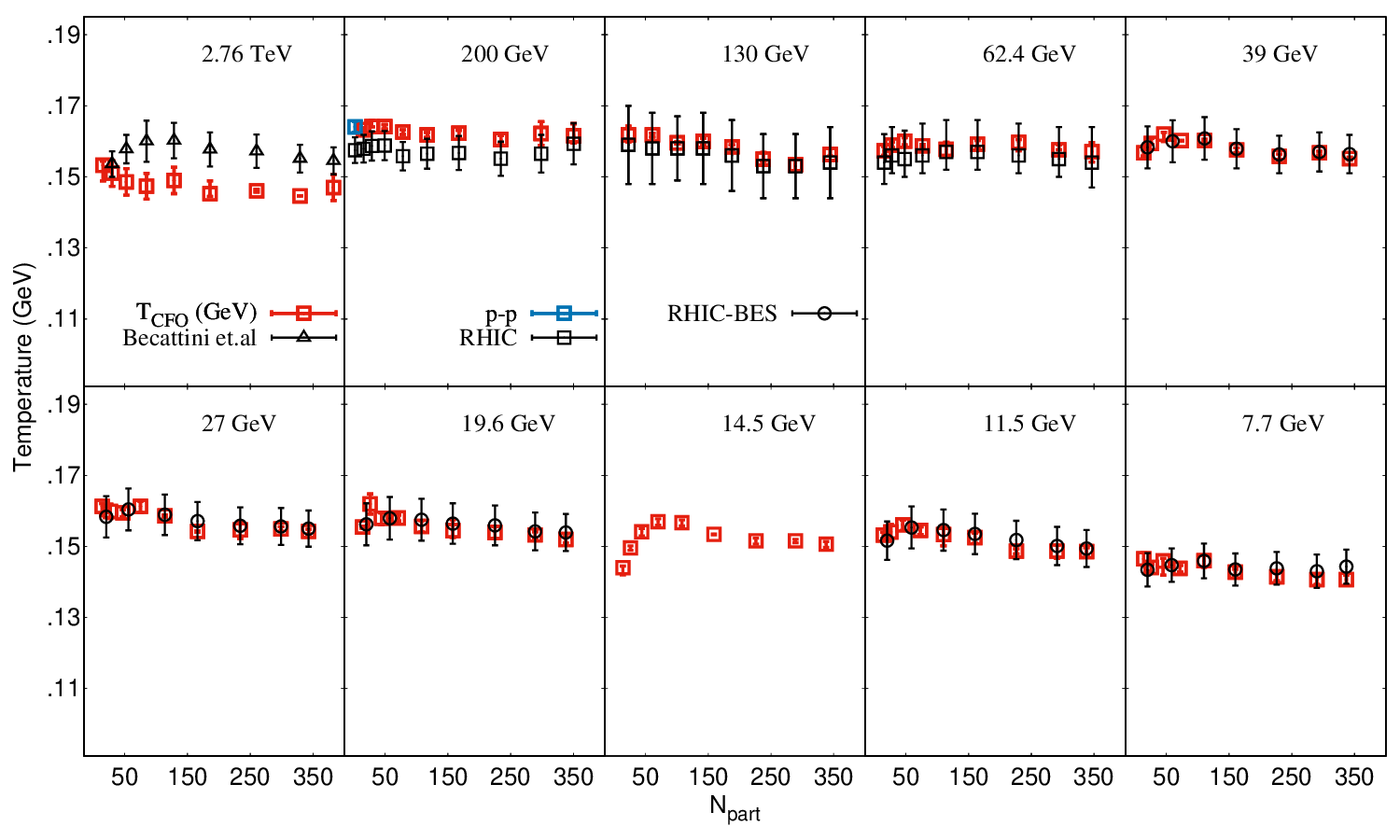}}
\label{fg.Ta}
}\\
\subfloat[]{
\hspace{-2em}
{\includegraphics[width=\textwidth]{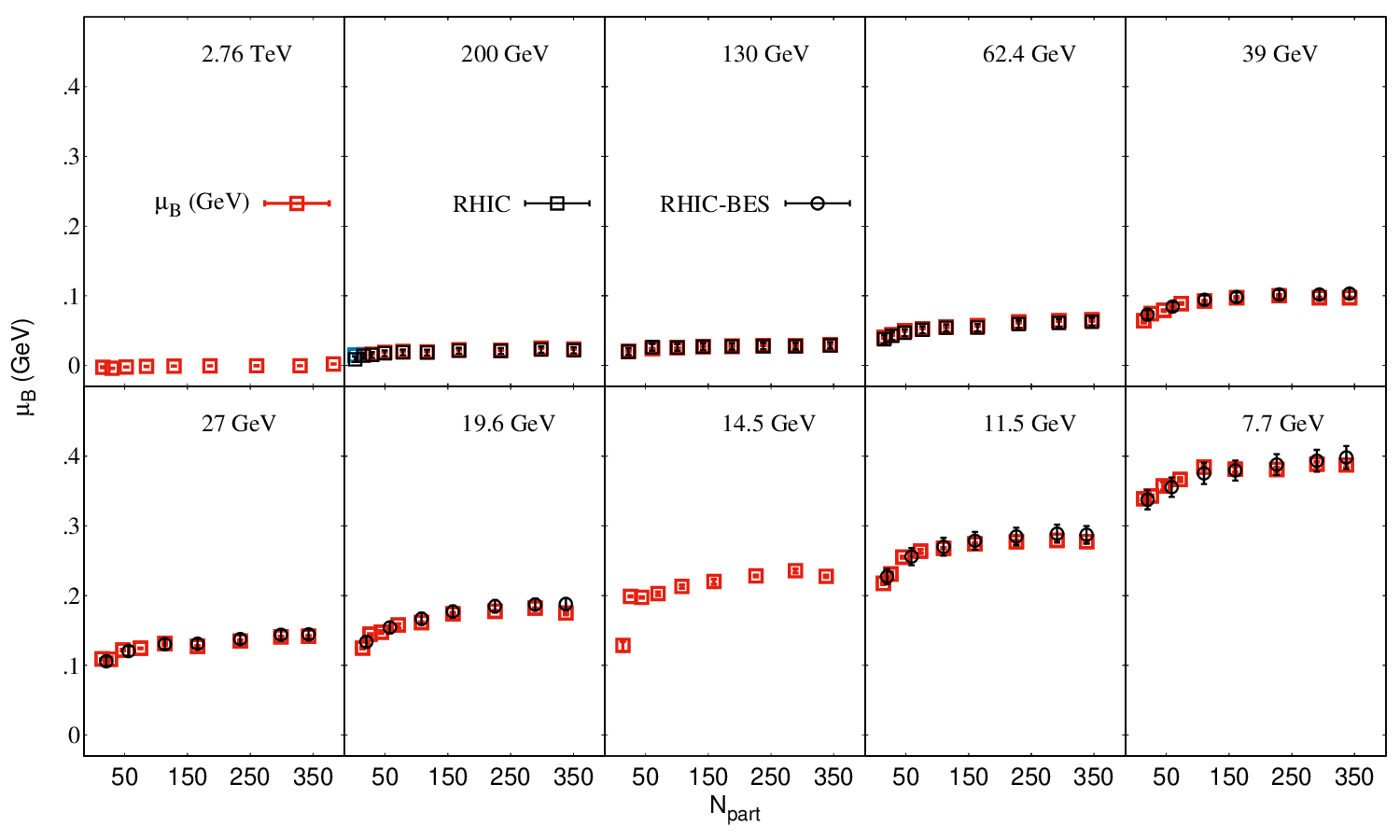}}
\label{fg.muB}
}\\
\caption{ (Color online) Variation of $T_{CFO}$, $\mu_B$ with $N_{part}$ for representative collision energies. Each column stands for different collision energy, ranging from LHC (2.76 TeV) to RHIC-BES (7.7 GeV). Most central collision ($0\%-5\%$) is denoted by highest value of $N_{part}$, lowest value denotes the peripheral collision ($70\%-80\%$). Red squares denote results for Pb-Pb at LHC and Au-Au at RHIC and BES. Blue square points analysis for p-p collision of RHIC-200 GeV. Results from available literature have been included following Ref.\cite{Becattini:2014hla} for LHC (black triangle), Ref.\cite{Abelev:2008ab} for RHIC (black square) and Ref.\cite{Adamczyk:2017iwn} for RHIC-BES (black circle). }
\end{figure*}

\begin{figure*}[hbt!]
\centering 
\subfloat[]{
\hspace{-2em}
{\includegraphics[width=\textwidth]{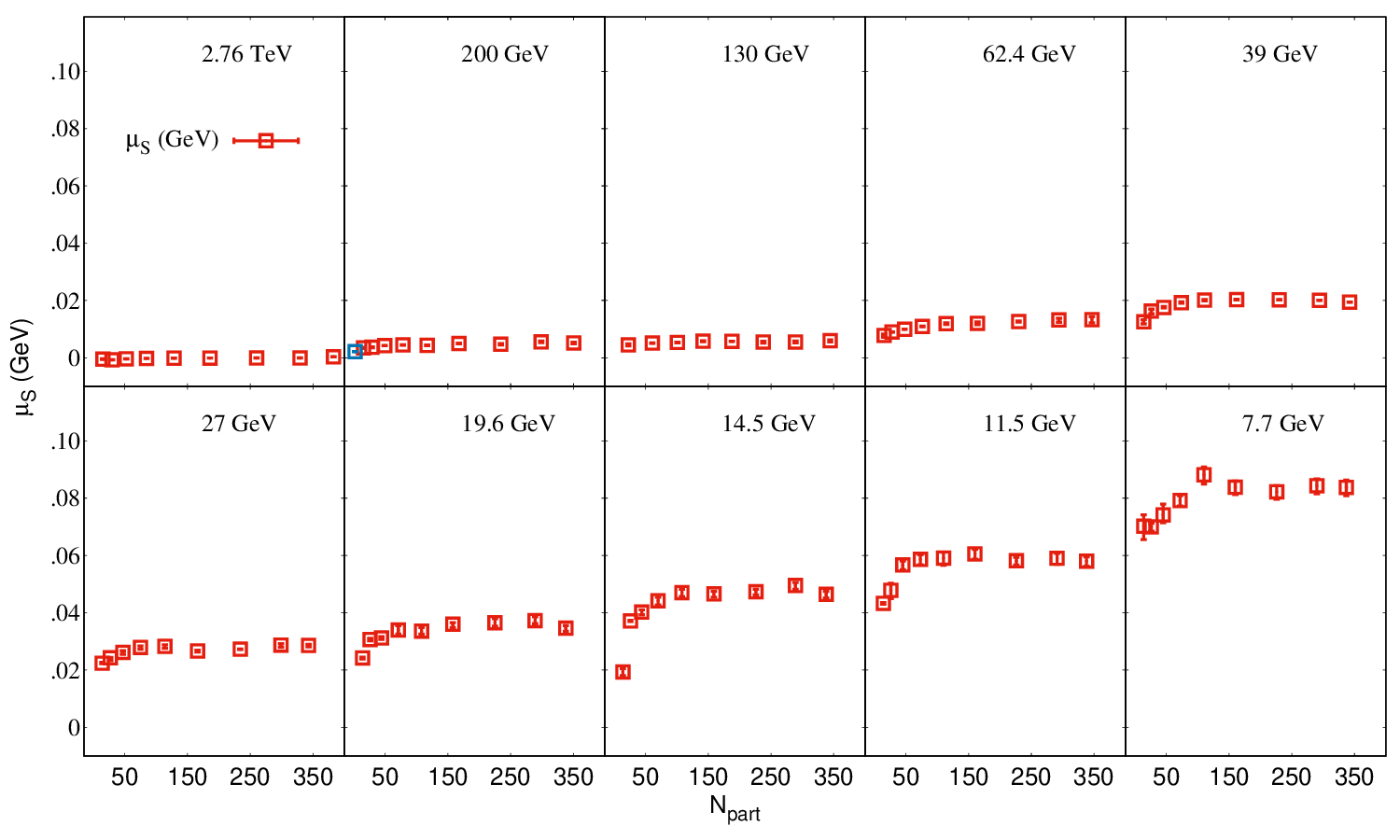}}
\label{fg.muS}
}\\
\subfloat[]{
\hspace{-2em}
{\includegraphics[width=\textwidth]{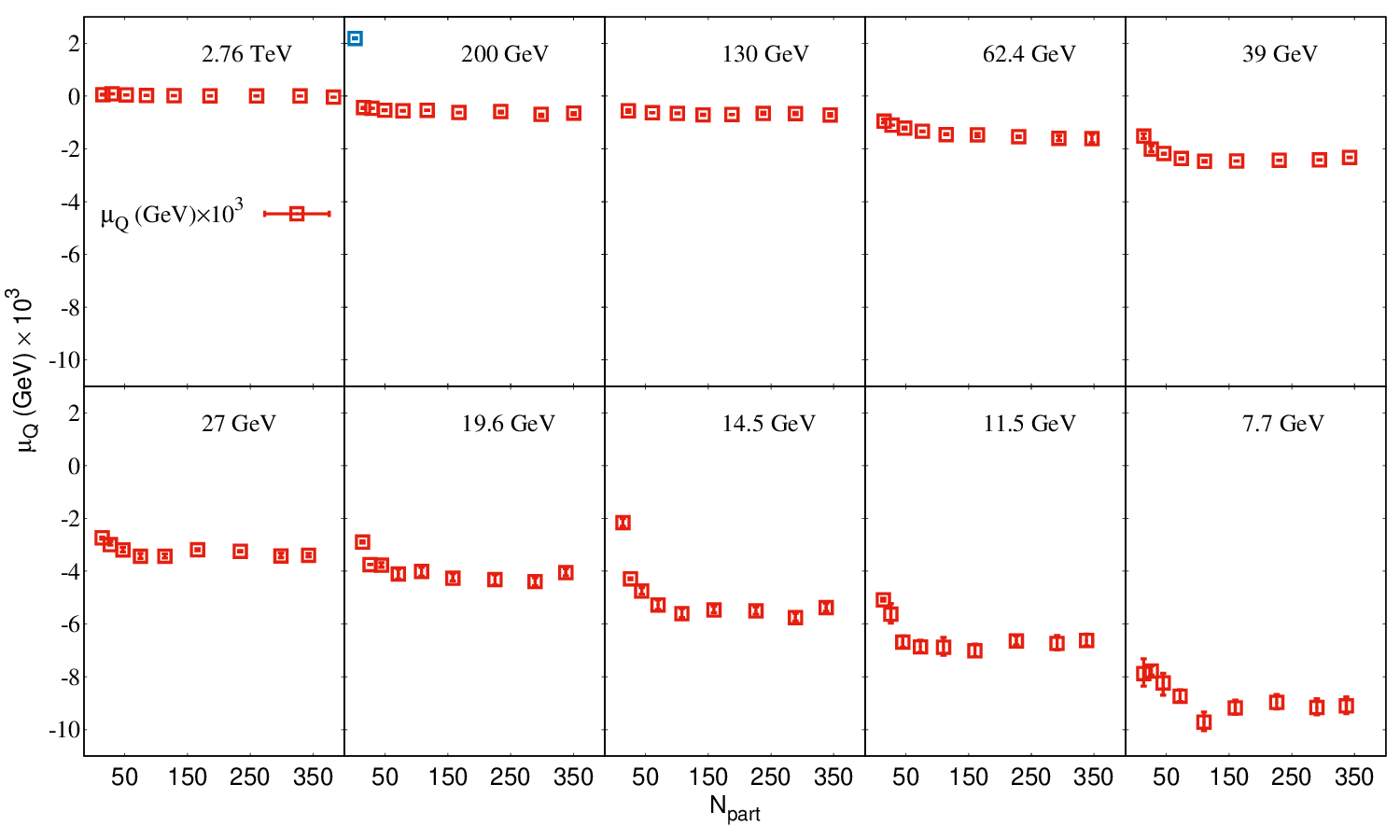}}
\label{fg.muQ}
}\\
\caption{ (Color online) Variation of $\mu_S$, $\mu_Q$ with $N_{part}$ for representative collision energies. The red square denotes results for Pb-Pb at LHC and Au-Au at RHIC and BES. Blue square points analysis for the p-p collision of RHIC-200 GeV. }
\end{figure*}

\begin{figure*}[hbt!]
\centering
\hspace{-2em}
{\includegraphics[width=\textwidth]{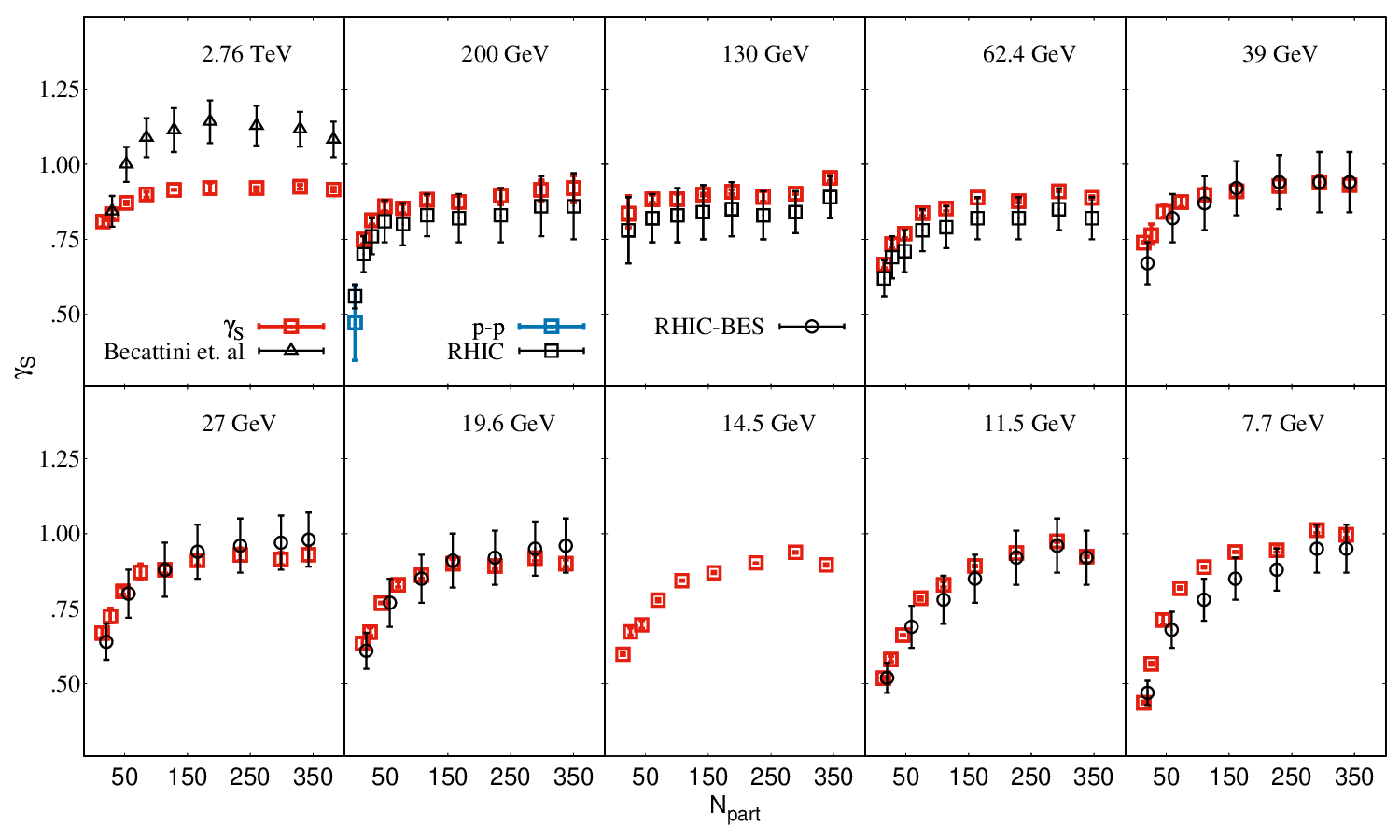}}
\caption{ (Color online) Variation of $\gamma_S$ with $N_{part}$ for representative collision energies. Each column stands for different collision energy, ranging from LHC (2.76 TeV) to RHIC-BES (7.7 GeV). Most central collision ($0\%-5\%$) is denoted by highest value of $N_{part}$, lowest value denotes the peripheral collision ($70\%-80\%$). Red square denotes results for Pb-Pb at LHC and Au-Au at RHIC and BES. Blue square points analysis for p-p collision of RHIC-200 GeV. Results from available literature have been included following Ref.\cite{Becattini:2014hla} for LHC (black triangle), Ref.\cite{Abelev:2008ab} for RHIC (black square) and Ref.\cite{Adamczyk:2017iwn} for RHIC-BES (black circle). }
\label{fg.gammas}
\end{figure*}

We have described the variation of our extracted freeze-out parameters with the number of participants($N_{part}$) for various collision energies in Fig.[\ref{fg.Ta}, \ref{fg.muB}, \ref{fg.muS}, \ref{fg.muQ}, \ref{fg.gammas}]. In plots, the horizontal axis is the number of participants. Results for collision energy LHC$-$2.76 TeV to RHIC-BES$-$7.7 GeV have been shown in different columns in descending order from left to right. For the clarity of discussion, we shall discuss variation concerning $\sqrt{s_{NN}}$  first and then try to understand the changes with centrality. For completeness, we have also presented available results for $T_{CFO}, \mu_B$ and $\gamma_S$ from other studies alongside our findings. We have included results from Ref.\cite{Abelev:2008ab, Adamczyk:2017iwn} for RHIC, RHIC-BES and Ref.\cite{Becattini:2014hla} for LHC.  

The variation of the chemical freeze-out temperatures ($T_{CFO}$) in Fig.~[\ref{fg.Ta}] has good agreement with general understanding~\cite{Bhattacharyya:2019wag}. At most central collisions, the freeze-out temperature increases with collision energy, and near $\sqrt{s_{NN}}=39$ GeV saturates around the value of 160 MeV as it reaches the Hagedorn limit \cite{Hagedorn:1965st}. At LHC, $T_{CFO}$ is lower than the expected value as the proton yield is lower than their preceding RHIC energies \cite{Abelev:2013vea}. The freeze-out temperatures seem to have a weaker dependence on the centrality and appear to be independent of $N_{part}$ and maintains a flat pattern at all collision energies. For $\sqrt{s_{NN}}=14.5$ GeV, the variation of $T_{CFO}$ is a little different at peripheral bins. The horn-like behavior is arising from a relatively lower yield of the proton (evident from the ratio $\bar{p}/p$ and $p/\pi^+$), which also reflects in the extracted values of $\mu_B$. There is good agreement with other results (black points) for $T_{CFO}$. In Ref.\cite{Abelev:2008ab, Adamczyk:2017iwn}, the parametrization has been performed utilizing only pion, kaon, and proton. Results from these analyses match with our findings, whereas the little differences in LHC energy may arise from the fitting procedure and particle species used for fitting.

We have plotted the baryon chemical potential as a functions of $N_{part}$ for all $\sqrt{s_{NN}}$ in Fig.[\ref{fg.muB}]. The general expectation is that at lower collision energies, a larger amount of nucleons deposit in the collision region due to the baryon stopping \cite{Sorge:1995dp, E-802:1998xum}. But at very high collision energies, the nuclei are transparent to each other \cite{Gaardhoje:2004mc, Kapusta:2018omb}. So at higher RHIC and LHC energies, the medium is created having almost zero net baryon number. Therefore the net baryon density and hence the estimated chemical potential $\mu_B$ would decrease with increasing $\sqrt{s_{NN}}$ due to baryon transparency \cite{Kapusta:2018omb}. In the same manner, one should expect a rise of $\mu_B$ for higher $N_{part}$. In central collisions, the value of the deposited net baryon number increases due to baryon stopping among a large number of the participating nucleons. Contrarily for a peripheral collision, a lesser number of nucleons get deposited in the collision zone, creating a dilute system of net baryon, which results in a smaller value of $\mu_B$. We have observed this trend in all $\sqrt{s_{NN}}$. Our resulted $\mu_B$ agrees with previous findings from Ref.\cite{Abelev:2008ab, Adamczyk:2017iwn}.


Strangeness chemical potential $\mu_S$ shows a similar trend as $\mu_B$ in Fig.[\ref{fg.muS}]. It decreases as collision energy increases and becomes zero at LHC energy. On the other hand, $\mu_S$ escalates as one goes from peripheral to the central collision. The correlation between $\mu_S$ and $\mu_B$ can be described in the following manner. A higher baryon density demands hyperons to be produced more than anti-hyperons. To maintain the strange neutrality, this excess amount of strangeness from the baryonic sector has to be nullified from the mesonic sector. So in the mesonic sector, $K^+$ is more abundant than $K^-$. Being the lightest strange particles, this difference between charged kaons determines the sign and trend of $\mu_S$.

The general trend of charge chemical potential $\mu_Q$ is the same as other $\mu$s except for the sign. As $N_{part}$ increases, it becomes more negative and the magnitude decreases with $\sqrt{s_{NN}}$ in  Fig.[\ref{fg.muQ}]. However, $\mu_Q$ is more negative for larger baryon densities. We can understand this as following. The neutrons are more abundant than protons in the colliding heavy nuclei. This abundance generates a net negative isospin value in the collision system and produces more $\pi^-$ than $\pi^+$, to conserve the isospin. As the lightest charged particle, these pions determine the negative $\mu_Q$. This reasoning will be more clear if we look into the value of $\mu_Q$ for $\sqrt{s_{NN}}=200$  GeV at $N_{part}=2$. In this case of $p$-$p$ collision, the isospin dominance should not act in favor of $\pi^-$, as there is no neutron in the colliding particles. So one should expect the $\mu_Q$ to be positive for this case. Indeed we have observed a positive value of $\mu_Q$ for the $p$-$p$ collision of 200 GeV RHIC energy. The net value of isospin increases with the $N_{part}$, thus increases asymmetry between the yield of charged pions. So the magnitude of $\mu_Q$ rises following the $\mu_B$.
 
If strange particles achieve chemical equilibrium, then the thermal abundance of kaons should be described by equilibrium thermal parameters (T, $\mu$s) of a grand canonical ensemble. This is not observed in cases of a small collision system like $p$-$p$, $p$-$A$ and even in $A$-$A$ with a smaller $N_{part}$ \cite{Manninen:2008mg}. Several models \cite{Becattini:1997ii, Cleymans:1997ib, Hamieh:2000tk, Cleymans:2001at, Becattini:2003wp} have tried to describe this source of strangeness undersaturation in the smaller system and advocated the use of $\gamma_S$. The common perception from all this work is that $\gamma_{S}$ scales the deviation of strange particles from their respective equilibrium thermal yield of a grand canonical ensemble, while $\gamma_{S}=1$ denotes the equilibration in the strange sector. 

We have shown the variation of $\gamma_S$ in Fig.[\ref{fg.gammas}]. It is interesting to notice that even in LHC and high RHIC energies, the $\gamma_S$ has an increasing trend from lower peripheral to a central collision, though the temperature and other chemical potentials do not change much. Initially, it starts from a lower value in the case of peripheral collisions and increases with participants. Around $N_{part}=150$, $\gamma_S$ tends to saturate to the most central values. It appears that the strangeness tends to be closer to the grand canonical limit as the system size increases. The saturation of $\gamma_S$ with the colliding system size for central $A$-$A$ collisions suggests that the strangeness suppression may be independent of the hadronic scatterings, which happens in the later time of the evolution \cite{Becattini:1997ii}. From the pattern, one can also conclude that the strangeness equilibration has a prominent dependence on the number of participants and system volume. The differences between the values between peripheral and central are larger for lower RHIC-BES energies. The general understanding from the above study is that the strangeness sector may be closer to equilibrium in a peripheral collision of higher $\sqrt{s_{NN}}$ whereas the deviation from equilibrium is larger for peripheral cases in lower collision energy.  We have found that strangeness equilibration tends to happen in collisions with a $N_{part}$ more than $150$, which may be the threshold $N_{part}$ for the creation of a deconfined phase in a $A$-$A$ collision, which drives the system close to strangeness equilibration \cite{Koch:1986ud}. 

We want to mention that even at high RHIC and LHC energies, the central value of $\gamma_S$ lies below $1$ (around 0.9). This finding is in agreement with previous analyses from RHIC collaboration \cite{Abelev:2008ab, Adamczyk:2017iwn}. In Ref. \cite{Adamczyk:2017iwn}, the $\gamma_S$ is shown to increase and saturate near $1$ as more hyperon species are included in chemical freeze-out parametrization. In this context, our method would be similar to the standard chi-square analysis, where the parameterization depends on the chosen hadronic ratios. Future studies with other heavy ions and data of hyperons may help to understand this.
\subsection{Scaling Nature of CFO parameters}
\begin{figure*}[hbt!]
\centering
\hspace{-2em}
{\includegraphics[width=\textwidth]{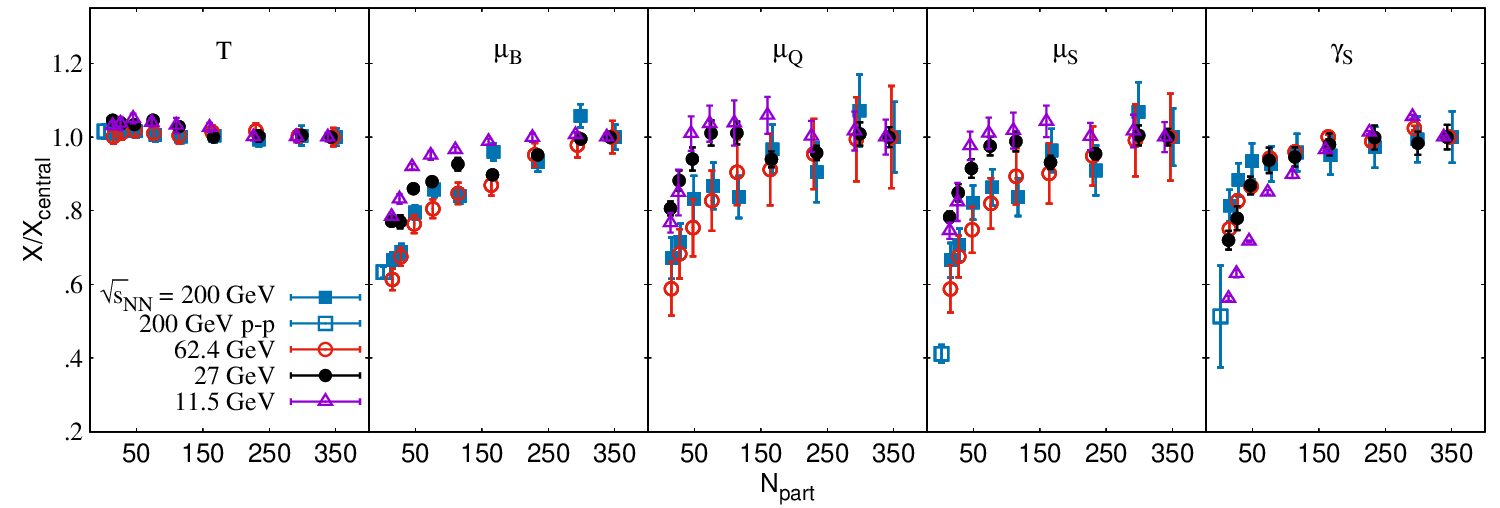}}
\caption{(Color online) Scaling behavior of various parameters with $N_{part}$. Representative points are for collision energies RHIC(200 GeV, 62.4 GeV) and RHIC-BES(27 GeV, 11.5 GeV). Different columns denote different scaled parameters.}
\label{fg.scale}
\end{figure*}
The scaling behavior with $N_{part}$ is important to calibrate the chemical composition at freeze-out with system size. To simplify the discussion, we have normalized the obtained parameters by their corresponding value for the most central collision of individual $\sqrt{s_{NN}}$. As an example, to understand the scaling of temperature at 200 GeV, we have divided the extracted $T_{CFO}$ of each centrality bin ($N_{part}$) with that of the most central collision (maximum $N_{part}$). Fig.[\ref{fg.scale}] shows the variation for all five freeze-out parameters. For simplicity, we have plotted for two collision energies from both RHIC (200, 62.4) and RHIC-BES (27, 11.5). These scaled quantities should lie around $1$ if parameters do not vary much with centrality ($N_{part}$). The scaled freeze-out temperature shows this pattern for all collision energies. It seems that for given incident energy, the freeze-out temperature does not vary much with the system size, whereas the scaled baryon chemical potential ($\mu_B$) has an increasing trend as it becomes maximum at most central collisions. In the case of equilibrium among all the charges, all the chemical potentials should commensurate with each other. Scaled $\mu_Q$ and $\mu_S$ should follow the pattern of $\mu_B$ with both the number of participants and collision energy, which we have already discussed in section [\ref{sc.freeze}].  We have indeed observed a similar trend for all three $\mu$s. There is a trend of saturation near $1$ around $N_{part}=150$. Onward this point, the system may achieve a thermodynamical state which is independent of the system size. Future analysis with other colliding ions at these c.m energies may shed light on this issue. Non-triviality could have appeared in the case of $\gamma_S$ as it is a non-equilibrium parameter. But the observed trend is similar to the chemical potentials. It starts from a smaller magnitude and saturates onward $N_{part}=150$. The system may have enough energy and number density for strangeness equilibration onward this centrality bin \cite{Kurkela:2018xxd}, and we can employ a grand canonical description to describe the yield at freeze-out. The deviation of scaled $\gamma_S$ from central value is larger for peripheral collisions in lower RHIC-BES energy, which indicates that colliding energy has a crucial contribution to decide the equilibration of strangeness.

\subsection{Particle Yield Ratios}
\label{sc.yratio}
In this section, we shall discuss ratios regarding detected particles to check the efficiency of our parameterization. We have estimated particle ratios from our extracted freeze-out parameters and plotted them alongside their experimental values. Variances in the detected yield ratios have been obtained using the standard error propagation method \cite{knoll2000radiation}, considering both the systematic and statistical uncertainties of data. We have calculated the variance of thermally estimated ratios by evaluating them at the extrema of the obtained freeze-out parameters.

\begin{figure*}[hbt!]
\centering
\hspace{-2em}
\subfloat[]{
{\includegraphics[width=\textwidth]{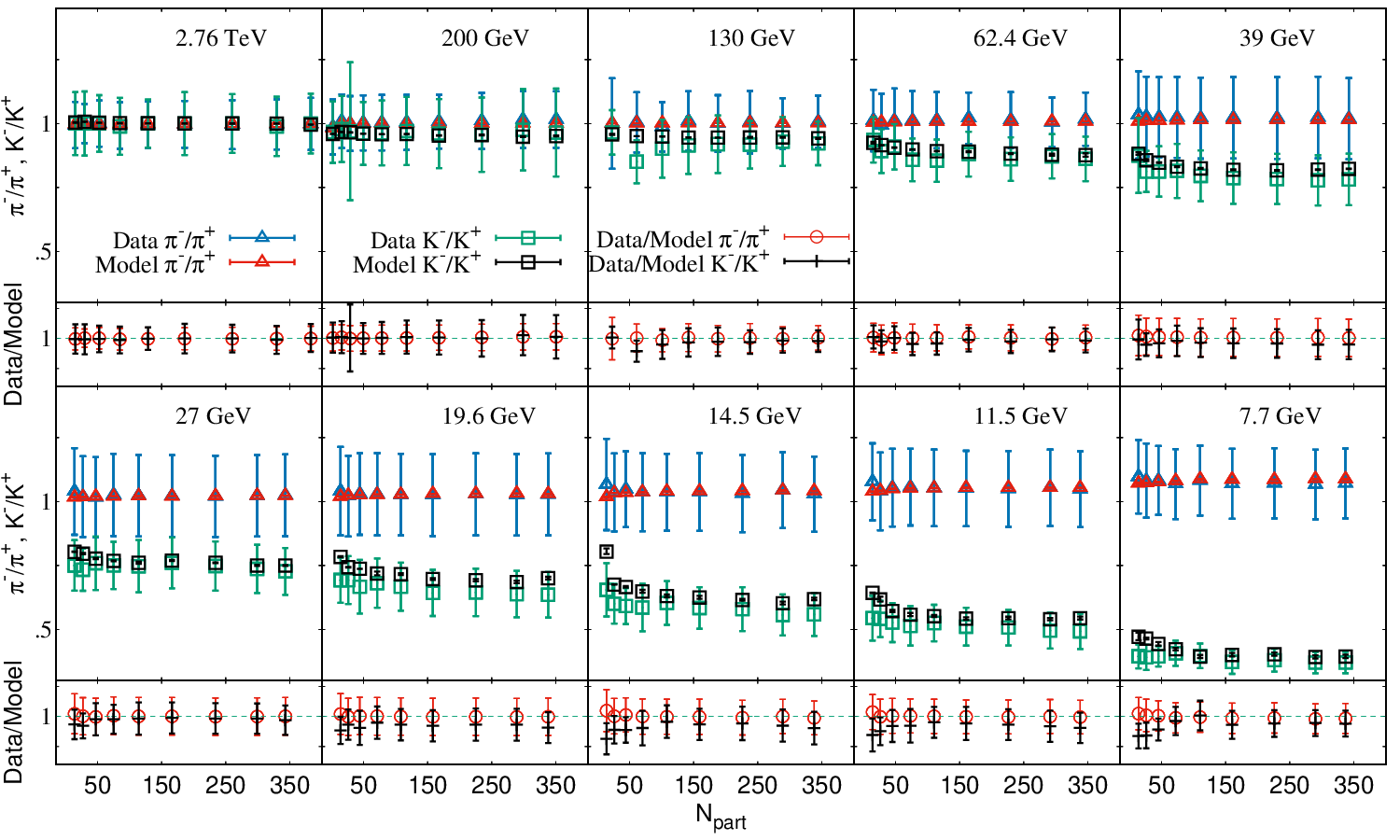}}
\label{fg.meson}
}\\
\hspace{-2em}
\subfloat[]{
{\includegraphics[width=\textwidth]{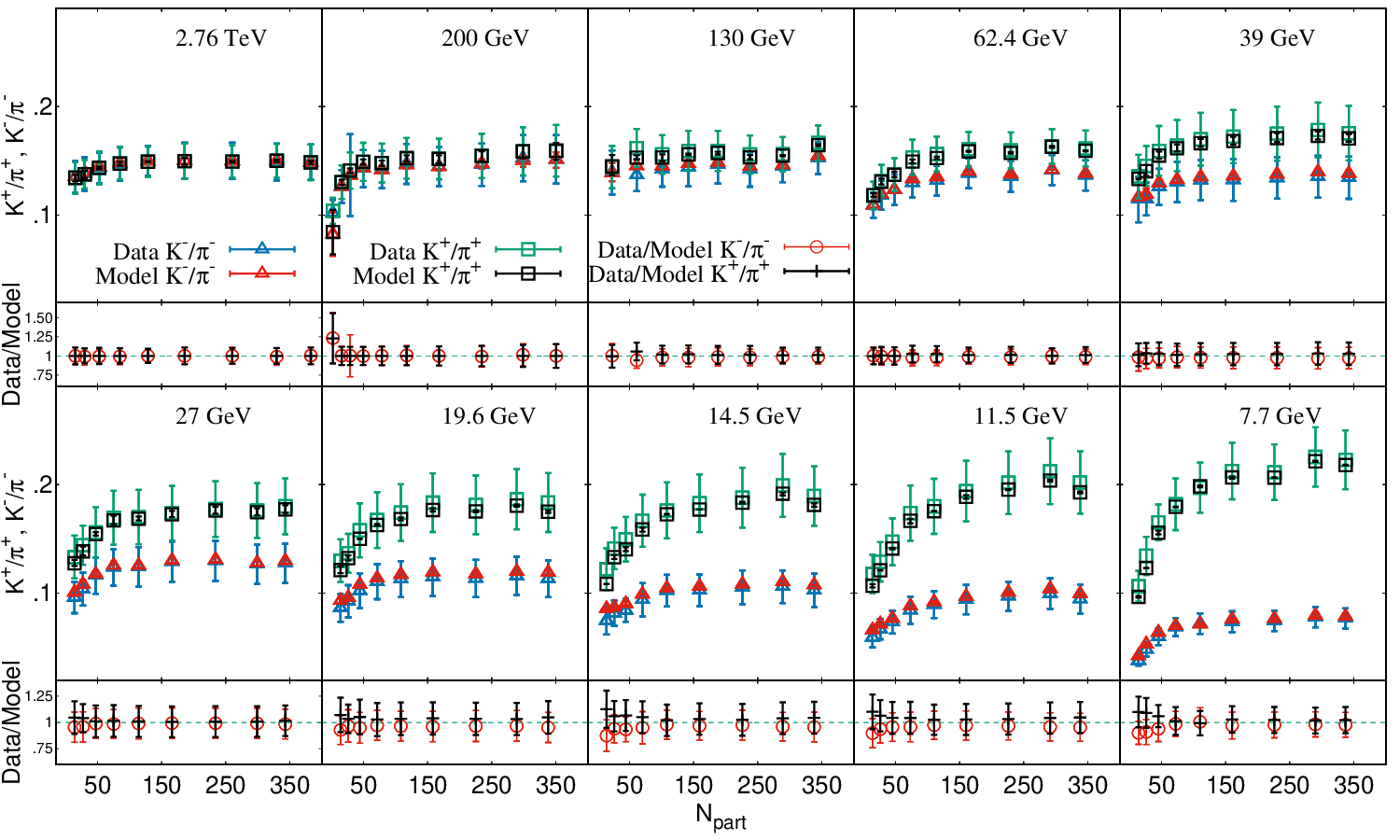}}
\label{fg.kpi}
}\\
\caption{(Color online) $N_{part}$ dependency of $\pi^-/\pi^+$, $K^-/K^+$ (upper panel) and 
$K^+/\pi^+$, $K^-/\pi^-$ (lower panel) for different collision energies ($\sqrt{s_{NN}}$). Experimental data (Blue and Green) are from RHIC~\cite{Abelev:2008ab}, RHIC-BES~\cite{Adamczyk:2017iwn, Adam:2019dkq} and LHC~\cite{Abelev:2013vea}. Model estimations are calculated from freeze-out parametrization. Ratio between the data and model are given below each plot.}
\label{fg.mkpi}
\end{figure*}
In Fig.[\ref{fg.meson}] we have plotted the particle to anti-particle ratios for pions and kaons. There is good agreement between model estimation and experimental data for both ratios. No notable variation has been observed for $\pi^-/\pi^+$ with $N_{part}$ and $\sqrt{s_{NN}}$. Freeze-out temperature and $\mu_Q$ determine the chemical abundance of pions. For the ratio of negatively charged to positively charged pions, the variation should depend on $\mu_Q$ only. Here we want to mention that there is no prominent variation of $\mu_Q/T_{CFO}$ with $\sqrt{s_{NN}}$ and $N_{part}$, as the value of $\mu_Q$ is much smaller (around $5~MeV$) than the value of $T_{CFO}$(about $150~MeV$). On the other hand, $\mu_Q$ is much lower than the mass of the pion itself. So it does not differentiate between the thermal yield of $\pi^-$ and $\pi^+$, and the ratio lies near unity for all collision energies and centrality classes.
         
The asymmetry between $K^-$ and $K^+$ depends on net baryon density and net strangeness from the hyperon sector. One should expect a larger yield of $K^+$ than $K^-$ at higher baryon density($\mu_B$). We have observed this pattern with both centrality and $\sqrt{s_{NN}}$. At lower RHIC-BES energies, the ratio is far from unity due to higher net baryon density and approaches $1$ as the collision energy increases. The yields of particle and anti-particle become equal at LHC as baryon transparency takes over, and the medium starts with zero net baryon density. With $N_{part}$, a commensurable trend has been observed following the value of $\mu_B$. At peripheral bins of lower collision energies, our model has overestimated the $K^-/K^+$ ratio. This overestimation has occurred as an interplay between the $\bar{p}/p$ ratio and the constraint $NetS=0$. The $\mu_B$ decreases towards peripheral and non-central collision following the $\bar{p}/p$ ratio, which results in a lower density of $\Lambda$ and other hyperons. On the other hand, to maintain the strangeness neutrality, thermal parameters adjust to produce a larger thermal density of $K^-$, which results in this overestimation.  

The charge independent $K/\pi$ ratios $K^+/\pi^+$, $K^-/\pi^-$ are important observables for understanding the strangeness production in high energy collisions. As the lightest mass hadrons, pions may act as the proxy of entropy, whereas the kaons carry the signature of strangeness. Strange particles are important in studying chemical equilibrium in heavy-ion collisions due to their late production \cite{Koch:1986ud}. A charged particle multiplicity ($dN_{ch}/d\eta$) dependent saturation of strangeness normalized to pions has already been observed in LHC \cite{ALICE:2017jyt}, which can be utilized to investigate the system size dependent strangeness production. Ref.\cite{Kurkela:2018xxd} has related this saturation to equilibration with a threshold $dN_{ch}/d\eta$. In heavy nuclei collisions,  the overlap region and $dN_{ch}/d\eta$ both are related to $N_{part}$ \cite{Abelev:2008ab}. We have observed the same saturation trend with $N_{part}$ here for all $\sqrt{s_{NN}}$ in Fig.[\ref{fg.kpi}] and have suitably reproduced it with our parametrization. This saturation starts around $N_{part}=150$ in higher RHIC and LHC. Here we want to mention that there is no variation of ${\pi^-}/{\pi^+}$ with centrality and collision energy, but $K^-/K^+$ has a strict dependence on both. At lower collision energy, $K^+/\pi^+$  is much higher than $K^-/\pi^-$ due to the excess yield of $K^+$. The difference between the ratios decreases with increasing collision energies, and they become equal at LHC, as the particle-antiparticle yields become the same. The pattern of $\gamma_S$ has a close resemblance to both the ratios. It seems that as $N_{part}$ decreases, the kaon yields deviate far from their equilibrium yield. So a non-equilibrium parameter $\gamma_S$ had to be introduced in our thermal model. Lower the value of $\gamma_S$, larger is the deviation from equilibrium for kaons.

Here we want to reemphasize that both $K^-/K^+$ and $\pi^-/\pi^+$ have no significant variation with centrality bins at LHC. This symmetry between particle and anti-particle demands $\mu_Q$ and $\mu_S$ to be almost zero. A centrality variation in the $K/\pi$ ratio cannot be reproduced with zero $\mu$ without introducing a $\gamma_S$ like parameters. This centrality variation of the $K/\pi$ ratio indicates that the strange particles are out of equilibrium at peripheral collisions of LHC.
\begin{figure*}[hbt!]
\centering
\hspace{-2em}
\subfloat[]{
{\includegraphics[width=\textwidth]{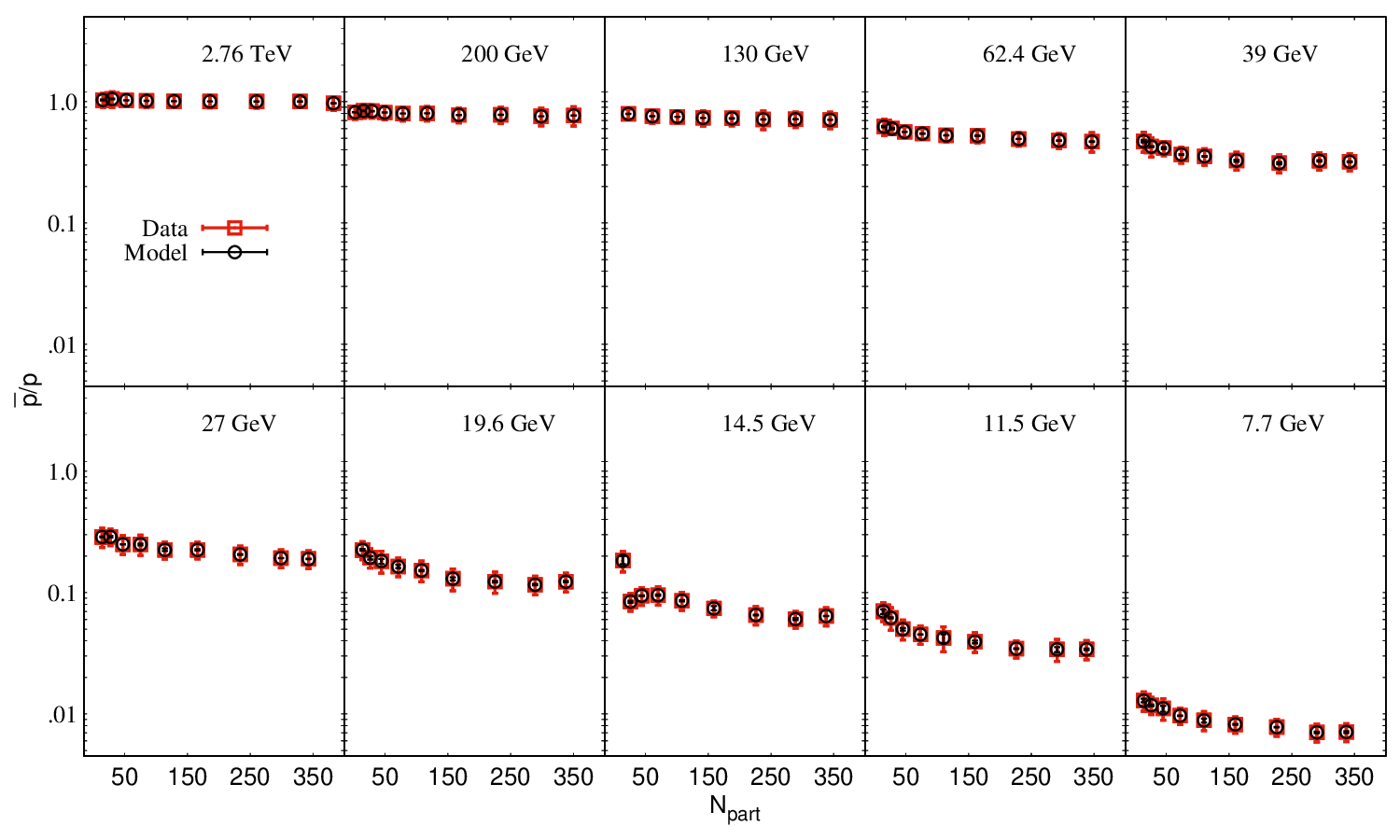}}
\label{fg.proton}
}\\
\hspace{-2em}
\subfloat[]{
{\includegraphics[width=\textwidth]{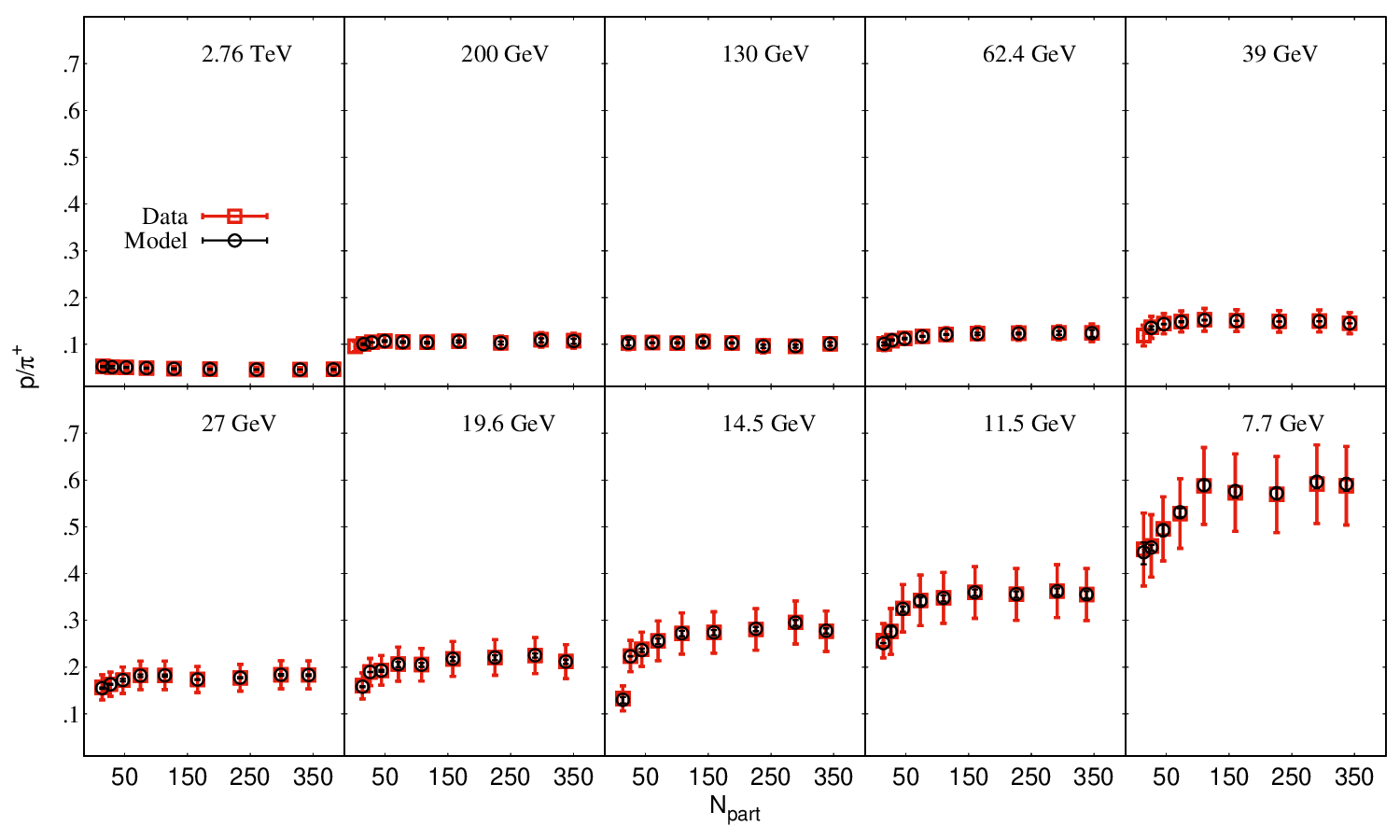}}
\label{fg.ppi}
}\\
\caption{(Color online) $N_{part}$ dependency of $\bar{p}/p$ (upper panel) and $p/\pi^+$ (lower panel) for different $\sqrt{s_{NN}}$. Experimental data (red) are from Ref.~\cite{Abelev:2008ab,  Abelev:2013vea, Adamczyk:2017iwn, Adam:2019dkq}. Thermal estimations (black) are estimated from freeze-out parametrization. }
\label{fg.baryon}
\end{figure*}
From the discussion of freeze-out parameters, it appears that the variation of anti-proton (anti baryon) to proton (baryon) is a guideline to understand the variation of $\mu_B$. This ratio becomes $1$ at upper RHIC and LHC energies as the colliding nuclei pass through each other and, the hadrons are created out of a medium having zero net baryon density. In lower collision energy, baryon stopping motivates a larger net baryon density. As a result, protons are more abundant than anti-proton and advocate a smaller $\bar{p}/p$ at lower collision energy. Our thermal model estimations have good agreement with the experimental data in Fig.[\ref{fg.proton}]. Towards peripheral collisions, this ratio tends to increase and symbolizes the decrease of baryon dominance over anti-baryons. Initial net baryon number density decreases as one goes from central to peripheral collision due to nuclear distribution \cite{Shou:2014eya} of the colliding nuclei and induces a smaller baryon anti-baryon asymmetry in their yield.
 
At this point, we also want to mention that for the ratio $\bar{p}/p$, there is a chance of over-fitting as our constructed ratio net-baryon to total baryon Eq.[\ref{eq.conserveb}] reduces to $\bar{p}/p$, as we have utilized centrality dependent data for (anti-)proton from the baryon sector. This deficiency of our formalism will reduce when centrality data for other (anti-) baryons are considered (as discussed in ref.\cite{Bhattacharyya:2019wag}). On that occasion, $\bar{p}/p$ will be an independent prediction.

We have discussed proton to positively charged pion ratio in Fig.~[\ref{fg.ppi}]. As we have already discussed, pions may act as the measure of entropy. So the ratio $p/\pi^+$ will describe the variation of baryon production with entropy. If the particles are produced only from deposited energy, then pions will be highly abundant than massive protons. But if the medium starts to evolve from a finite baryon density, then per pion, proton production will be larger to conserve the net baryon density. That is why a clear increasing trend for $p/\pi^+$ takes place as the collision energy decreases. This same increment is expected with $N_{part }$, as more baryon deposition happens in the more central collisions. This variation is prominent in lower $\sqrt{s_{NN}}$ due to the higher efficacy of baryon stopping.

\section{\label{sec:conclusion}Summary and conclusion} 
The equilibration of the system created in a heavy-ion collision should strongly depend on the system size and number of participants. A comparison among the chemical freeze-out conditions of high and low multiplicity $A$-$A$ collisions may shed light in that direction. Instead of the general $\chi^2$ minimization, a conserved charge-dependent parametrization process has been adopted, utilizing the mid-rapidity yield of the pion, kaon, proton to explore the freeze-out parameters of various centrality bins of $A$-$A$ collision for LHC (2.76 TeV), RHIC (200 GeV, 130 GeV, 62.4 GeV) and RHIC-BES (39 GeV, 27 GeV, 19.6 GeV, 14.5 GeV, 11.5 GeV, and 7.7 GeV). We have incorporated a strangeness suppression factor ($\gamma_S$) to estimate the possible non-equilibrium in strange hadrons in the peripheral collisions. We have discussed the variation of these chemical freeze-out parameters with both centrality and collision energy.

The variation of parameters with collision energies has good agreement with general understanding, whereas there are significant variations with the number of participants. The extracted freeze-out temperature has no strong dependence on the number of participants (centrality), whereas the chemical potentials show a wide variation. We have presented the behavior of scaled parameters to have a better understanding of the centrality variation. These parameters have been normalized with the value obtained in most central collisions and compared along with the other collision energies. Scaled $\mu_Q$ and $\mu_S$ appear to follow the scaling behavior of $\mu_B$, which may be a signal of equilibration among three conserved charges.  The strangeness suppression factor ($\gamma_S$) deviates from the equilibrium value at peripheral collisions and tends to saturate near unity in central collisions. In the peripheral collisions, $\gamma_S$ starts around $.7$ and increases towards most central bins. This variation indicates that the strange hadrons are deviated from equilibrium at low multiplicity collision, whereas there is a sign of equilibration as the $N_{part}$ increases. The flattening of the scaled parameter and $\gamma_S$ appears around a threshold of $N_{part}=150$. So we can apply a grand canonical description for the systems created out of $A$-$A$ collisions with more participants than $150$. We have found the $\gamma_S$ to lie below $1$ (around $0.9$) even at most central collisions. In this study, we have only used kaons, so the variation of $\gamma_S$ is an artifact of the kaon to pion ratio. Future analysis with yields data of other strange hyperons may help to understand this further.
  
Further, we have estimated different particle ratios to cross-check the effectiveness of our parameterization. Our estimated hadron ratios seem to have good agreement with experimental data. We have only reproduced ratios regarding pions, kaons, and protons as they are present in our analysis. A saturating trend with $N_{part}$ has been observed for the kaon to pion ratio and explained with the $\gamma_S$.
 
We want to mention that the centrality variation has previously been investigated in RHIC-BES energy \cite{Adamczyk:2017iwn}, and in LHC \cite{Becattini:2014hla, Sharma:2018jqf, Vovchenko:2019kes} with the $\chi^2$ approach. Rather than the conventional practice, we have followed a fitting procedure that relies on the conserved quantities and produces similar parameter sets. The agreement with other studies will act as a benchmark for the future application of this parameterization. Further, we have found a threshold $N_{part}$, which is significant to study the bulk properties in a thermodynamic picture. 
   
\section*{Acknowledgements}
This work is funded by UGC and DST. The author thanks Sumana Bhattacharyya, Sanjay K. Ghosh, and Rajarshi Ray for helpful discussions, and Pratik Goshal and Pracheta Singha for the critical reading of the manuscript. The publication and article processing charges are funded by SCOAP3.


\bibliography{refcentral}

\end{document}